\documentclass[aps,preprint,showpacs,preprintnumbers,amsmath,amssymb]{revtex4-2}
\usepackage{graphicx}
\usepackage{float}
\usepackage{epstopdf}
\begin{document}

\title{Z boson emission by a neutrino in de Sitter expanding universe}
\author{ Mihaela-Andreea B\u aloi, Cosmin Crucean, Diana Dumitrele\\ \thanks{mihaela.baloi88@e-uvt.ro, cosmin.crucean@e-uvt.ro, diana.dumitrele97@e-uvt.ro}\\
{\small \it Faculty of Physics, West University of Timi\c soara,}\\
{\small \it V. Parvan Avenue 4 RO-300223 Timi\c soara,  Romania}}

\begin{abstract}
Production of Z bosons in emission processes by neutrinos in the expanding de Sitter universe is studied by using perturbative methods. The total probability and transition rate for the spontaneous emission of a Z boson by a neutrino is computed analytically, then we perform a graphical analysis in terms of the expansion parameter. Our results prove that this process is possible only for large expansion conditions of the early Universe. Finally the density number of Z bosons is defined and we obtain a quantitative estimation of this quantity in terms of the density number of neutrinos.
\end{abstract}

\maketitle

\newpage
\section{Introduction}
The problem of electro-weak interactions in a de Sitter space-time by using perturbative methods was studied only recently in \cite{1,cd,cf}. In \cite{1,cd} the general formalism for studying the  neutral  current interactions intermediated by the Z boson was constructed in a curved space-time. This allows us to explore processes of interaction that generate production of massive bosons and fermions in an expanding de Sitter universe by adapting the electro-weak perturbation theory \cite{5,6,8,10,11,12,13,14,cr} to a curved space-time. It is well known that the massive bosons were produced in early universe \cite{10}, and it is important to explore all possible processes that could produce them, including the first order perturbative processes that are forbidden in Minkowski theory \cite{14} by energy and momentum conservation. In a non-stationary space-time the translational invariance with respect to time is lost and the amplitudes and probabilities corresponding to processes of spontaneous particle production are non-vanishing \cite{23,24}.

The idea of exploring the problem of particle generation at fields interactions was first proposed in \cite{19}, and the main results of this paper are related to the fact that the perturbative calculations can be translated in the number of particles. Another result established in \cite{19} is related to the conditions in which the perturbative particle production becomes dominant in rapport with the cosmological particle production.
In the present paper we want to study for the first time the process of Z boson emission by a neutrino in a de Sitter metric. We will use in our study the solutions of the Dirac equation and Proca equation in a de Sitter geometry \cite{3,4}, which have a defined momentum and helicity. We will use the perturnative formalism employed in the field theory including renormalization procedures for extracting finite results from our computations. Our approach allows us to explore the interesting limit cases when the expansion parameter is vanishing and the case when the expansion is large comparatively with the Z boson mass.

The paper begins in section two with the computations of the transition amplitude and probability for the process of Z boson emission by a neutrino. In section three we compute the total probability of the process and in section four the transition rate is obtained. Density number of Z bosons obtained in emission processes by neutrinos are analysed in section five and in Appendices we present the free fields solutions in de Sitter geometry and the integrals that help us to establish the analytical results. We use natural units with $\hbar=1,c=1$.

\section{Amplitude of Z boson emission by neutrino}
For analyse the emission of Z bosons by neutrinos in early universe we start with the de Sitter metric \cite{2}:
\begin{equation}\label{metr}
ds^2=dt^2-e^{2\omega t}d\vec{x}\,^2=\frac{1}{(\omega t_{c})^2}(dt_{c}^2-d\vec{x}\,^2),
\end{equation}
where the conformal time is given in terms of proper time by $t_{c}=\frac{-e^{-\omega t}}{\omega}$, and $\omega$ is the expansion factor ($\omega>0$).
The first order transition amplitude in electro-weak theory on curved space-time  for the interaction between Z bosons and the neutrino-antineutrino field was obtained in \cite{1}. For the process of Z emission by a single neutrino $\nu\rightarrow \nu+Z$, the transition amplitude is:
\begin{eqnarray}\label{am2}
\mathcal{A}_{\nu\rightarrow Z\nu}=-\int d^4x\sqrt{-g}\left(\frac{e_0}{\sin(2\theta_{W})}\right)(\overline{U}_{p',\sigma'})_{\nu}(x)\gamma^{\hat\mu}e_{\,\hat\mu}^{\alpha}\left(\frac{1-\gamma^{5}}{2}\right)
(U_{p\sigma})_{\nu}(x)f_{\alpha\mathcal{P},\lambda,Z}^*(x),
\end{eqnarray}
where $e_0$ is the electric charge, $\theta_{W}$ is the Weinberg angle and, $(U_{p\sigma})_{\nu}(x)$ is the solution of zero mass for the Dirac equation in de Sitter space-time \cite{3}, while $A_{\alpha}(Z)$ designates the Z boson field. The solutions for the free fields equations in momentum-helicity basis are presented in Appendix A. We also mention that we use point independent Dirac matrices $\gamma^{\hat\mu}$ and the tetrad fields $e_{\hat\mu}^{\alpha}$. For the
line element (\ref{metr}), in the Cartesian gauge, the tetrad components are:
\begin{equation}
e^{0}_{\widehat{0}}=-\omega t_c  ;\,\,\,e^{i}_{\widehat{j}}=-\delta^{i}_{\widehat{j}}\,\omega t_c.
\end{equation}
Our computations are done in the chart with conformal time $t_{c}\in(-\infty,0)$, which covers the expanding portion of de Sitter space.

\subsection{The calculation}
Using the solutions given in equations (\ref{sol1}), (\ref{sol2}) and (\ref{fo}) from Appendix A, the amplitude for longitudinal modes with $\lambda=0$, can be brought to the form:
\begin{eqnarray}\label{am0}
\mathcal{A}_{\nu\rightarrow Z\nu}(\lambda=0)=-\int d^4x\sqrt{-g}\left(\frac{e_0}{\sin(2\theta_{W})}\right)(\overline{U}_{p',\sigma'})_{\nu}(x)\gamma^{\hat0}e_{\,\hat0}^{0}\left(\frac{1-\gamma^{5}}{2}\right)
(U_{p\sigma})_{\nu}(x)f_{0\mathcal{P},\lambda=0,Z}^*(x)\nonumber\\
-\int d^4x\sqrt{-g}\left(\frac{e_0}{\sin(2\theta_{W})}\right)(\overline{U}_{p',\sigma'})_{\nu}(x)\gamma^{\hat i}e_{\,\hat i}^{j}\left(\frac{1-\gamma^{5}}{2}\right)
(U_{p\sigma})_{\nu}(x)f_{j\,\mathcal{P},\lambda=0,Z}^*(x).
\end{eqnarray}
In the case of transversal modes with $\lambda=\pm1$, only the spatial part of the solution gives contribution, since there are no temporal component of the solution i.e $f_{0\vec{\mathcal{P}},\lambda=\pm 1}(x)=0$, and we obtain:
\begin{eqnarray}\label{am1}
\mathcal{A}_{\nu\rightarrow Z\nu}(\lambda=\pm1)=-\int d^4x\sqrt{-g}\left(\frac{e_0}{\sin(2\theta_{W})}\right)(\overline{U}_{p',\sigma'})_{\nu}(x)\gamma^{\hat i}e_{\,\hat i}^{j}\left(\frac{1-\gamma^{5}}{2}\right)
(U_{p\sigma})_{\nu}(x)f_{j\,\mathcal{P},\lambda=\pm1,Z}^*(x).
\nonumber\\
\end{eqnarray}
The spatial integrals give the delta Dirac function expressing the momentum conservation in the emission process. For the temporal integral the new integration variable is $z=-t_{c}$ \cite{23}, and we use Bessel K functions by transforming the Hankel functions
$H^{(1,2)}_{\nu}(z)=\mp \left(\frac{2i}{\pi}\right)e^{\mp
i\pi\nu/2}K_{\nu}(\mp iz)$.
Then the amplitudes equations for $\lambda=0$ and $\lambda=\pm1$ are:
\begin{eqnarray}\label{l0}
&&\mathcal{A}_{\nu\rightarrow Z\nu}(\lambda=0)=\frac{e_0}{\sin(2\theta_{W})}\,\delta^3(\vec{\mathcal{P}}+\vec{p}\,'-\vec{p}\,)\frac{1}{\sqrt{\pi}(2\pi)^{3/2}}
\left(\frac{1}{2}-\sigma\right)\left(\frac{1}{2}-\sigma'\right)\nonumber\\
&&\times\biggl\{ \frac{\mathcal{P}\omega}{M_Z}A(t_{c})\xi^+_{\sigma'}(\vec{p}\,')\vec{\sigma}\cdot\vec{\epsilon}\,^*(\vec{n}_{\mathcal{P}},\lambda=0)\xi_{\sigma}(\vec{p}\,)
+\frac{\mathcal{P}\omega}{M_Z}C(t_{c})\xi^+_{\sigma}(\vec{p}\,')\xi_{\sigma}(\vec{p}\,)\biggl\},
\end{eqnarray}
\begin{eqnarray}\label{l1}
&&\mathcal{A}_{\nu\rightarrow Z\nu}(\lambda=\pm1)=\frac{e_0}{\sin(2\theta_{W})}\,\delta^3(\vec{\mathcal{P}}+\vec{p}\,'-\vec{p}\,)\frac{1}{\sqrt{\pi}(2\pi)^{3/2}}
\left(\frac{1}{2}-\sigma\right)\left(\frac{1}{2}-\sigma'\right)\nonumber\\
&&\times\biggl\{B(t_{c})\xi^+_{\sigma'}(\vec{p}\,')\vec{\sigma}\cdot\vec{\epsilon}\,^*(\vec{n}_{\mathcal{P}},\lambda=\pm1)\xi_{\sigma}(\vec{p}\,)\biggl\},
\end{eqnarray}
The notations $A(t_{c}),B(t_{c}),C(t_c)$ stand for the following temporal integrals:
\begin{eqnarray}\label{abc1}
A(t_{c})&=&\int_0^\infty dz \sqrt{z}\,e^{-i(p'-p)z}K_{-ik}(i\mathcal{P}z)\frac{1}{\mathcal{P}}\left(\frac{1}{2}-ik\right)-i\int_0^\infty dz z^{3/2}\,e^{-i(p'-p)z}K_{1-ik}(i\mathcal{P}z),\nonumber\\
C(t_c)&=&i\int_0^\infty dz z^{3/2}\,e^{-i(p'-p)z}K_{-ik}(i\mathcal{P}z),\nonumber\\
B(t_{c})&=&i\int_0^\infty dz \sqrt{z}\,e^{-i(p'-p)z}K_{-ik}(i\mathcal{P}z).
\end{eqnarray}

By using the results for the integrals with Bessel functions \cite{22} the final forms for the amplitudes are:
\begin{eqnarray}\label{affin}
&&\mathcal{A}_{\nu\rightarrow Z\nu}(\lambda=0)=\frac{e_0}{\sin(2\theta_{W})}\,\delta^3(\vec{\mathcal{P}}+\vec{p}\,'-\vec{p}\,)\frac{1}{(2\pi)^{3/2}}
\left(\frac{1}{2}-\sigma\right)\left(\frac{1}{2}-\sigma'\right)\nonumber\\
&&\times\biggl\{A_{k}(\mathcal{P},p,p')\xi^+_{\sigma'}(\vec{p}\,')\vec{\sigma}\cdot\vec{\epsilon}\,^*(\vec{n}_{\mathcal{P}},\lambda=0)\xi_{\sigma}(\vec{p}\,)\nonumber\\
&&+C_{k}(\mathcal{P},p,p')\xi^+_{\sigma'}(\vec{p}\,')\xi_{\sigma}(\vec{p}\,)\biggl\},\nonumber\\
&&\mathcal{A}_{\nu\rightarrow Z\nu}(\lambda=\pm1)=\frac{e_0}{\sin(2\theta_{W})}\,\delta^3(\vec{\mathcal{P}}+\vec{p}\,'+\vec{p}\,)\frac{1}{(2\pi)^{3/2}}
\left(\frac{1}{2}-\sigma\right)\left(\frac{1}{2}-\sigma'\right)\nonumber\\
&&\times\biggl\{B_{k}(\mathcal{P},p,p')\xi^+_{\sigma'}(\vec{p}\,')\vec{\sigma}\cdot\vec{\epsilon}\,^*(\vec{n}_{\mathcal{P}},\,\lambda=\pm1)\xi_{\sigma}(\vec{p}\,)\biggl\}.
\end{eqnarray}
The functions $A_{k}(\mathcal{P},p,p'),B_{k}(\mathcal{P},p,p'),C_{k}(\mathcal{P},p,p')$ that define the amplitudes are:
\begin{eqnarray}\label{ab}
&&A_{k}(\mathcal{P},p,p')=\frac{i^{-3/2}(2\mathcal{P})^{-ik}}{(\mathcal{P}+p\,'-p)^{3/2-ik}}\frac{\omega}{M_Z}
\Gamma\left(\frac{3}{2}-ik\right)\Gamma\left(\frac{3}{2}+ik\right)\left(\frac{1}{2}-ik\right)\nonumber\\
&&\times_{2}F_{1}\left(\frac{3}{2}-i
k,\frac{1}{2}-ik;2;\frac{p'-\mathcal{P}-p}{\mathcal{P}+p'-p}\right)
-\frac{i^{-3/2}\mathcal{P}(2\mathcal{P})^{1-ik}}{2(\mathcal{P}+p'-p)^{7/2-ik}}\frac{\omega}{M_Z}
\Gamma\left(\frac{7}{2}-ik\right)\nonumber\\
&&\times\Gamma\left(\frac{3}{2}+ik\right)\,_{2}F_{1}\left(\frac{7}{2}-i
k,\frac{3}{2}-ik;3;\frac{p'-\mathcal{P}-p}{\mathcal{P}+p'-p\,}\right),\nonumber\\
&&C_k(\mathcal{P},p,p')=\frac{i^{-3/2}\mathcal{P}(2\mathcal{P})^{-ik}}{2(\mathcal{P}+p'-p)^{5/2-ik}}\frac{\omega}{M_Z}
\Gamma\left(\frac{5}{2}-ik\right)\Gamma\left(\frac{5}{2}+ik\right)\nonumber\\
&&\times\,_{2}F_{1}\left(\frac{5}{2}-i
k,\frac{1}{2}-ik;3;\frac{p'-\mathcal{P}-p}{\mathcal{P}+p'-p\,}\right),
\end{eqnarray}
\begin{eqnarray}\label{cb}
&&B_{k}(\mathcal{P},p,p')=\frac{i^{-1/2}(2\mathcal{P})^{-ik}}{(\mathcal{P}+p'-p)^{3/2-ik}}
\Gamma\left(\frac{3}{2}-ik\right)\Gamma\left(\frac{3}{2}+ik\right)\nonumber\\
&&\times\,_{2}F_{1}\left(\frac{3}{2}-i
k,\frac{1}{2}-ik;2;\frac{p'-\mathcal{P}-p}{\mathcal{P}+p'-p\,}\right).
\nonumber\\
\end{eqnarray}

\begin{figure}[h!t]
\includegraphics[scale=0.35]{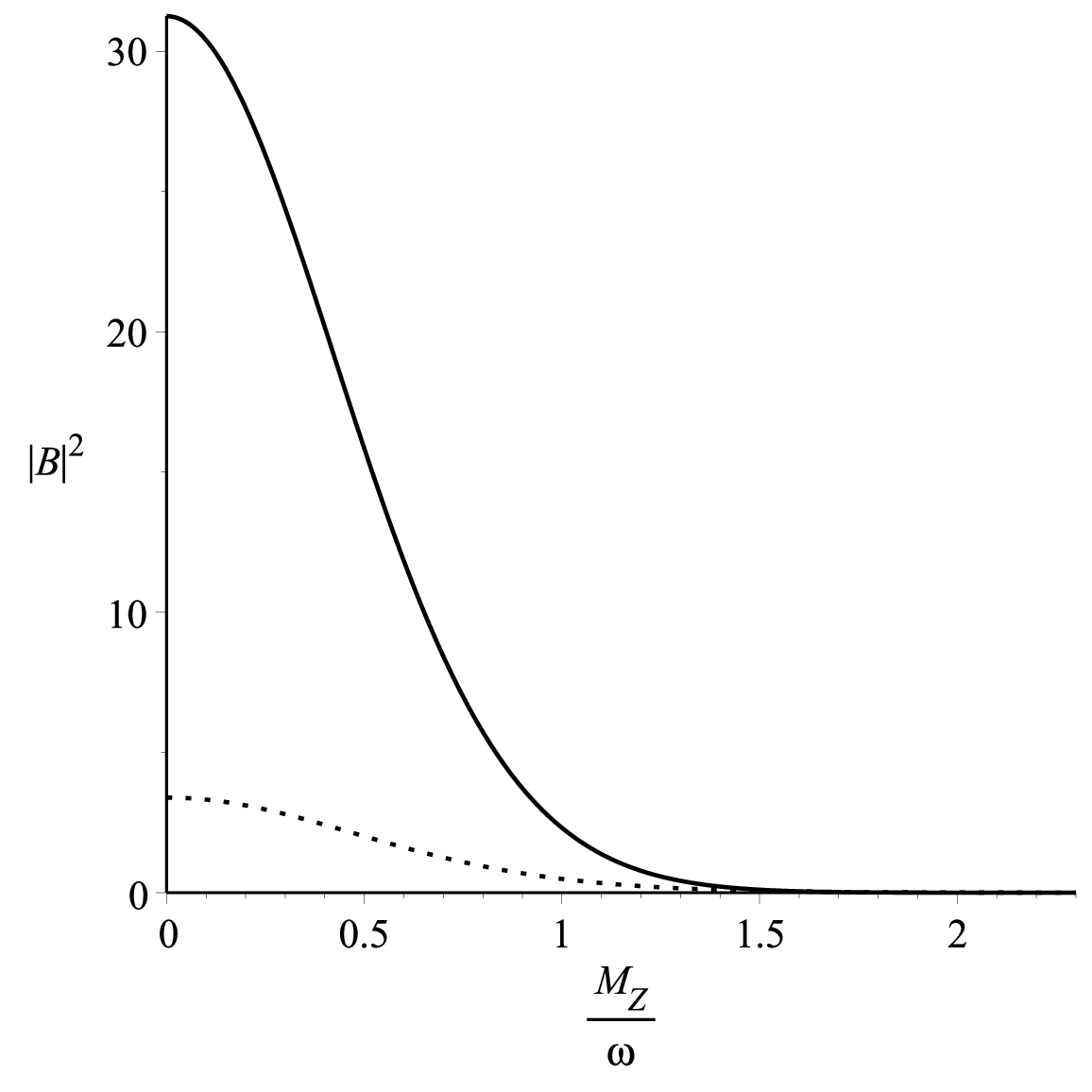}
\includegraphics[scale=0.35]{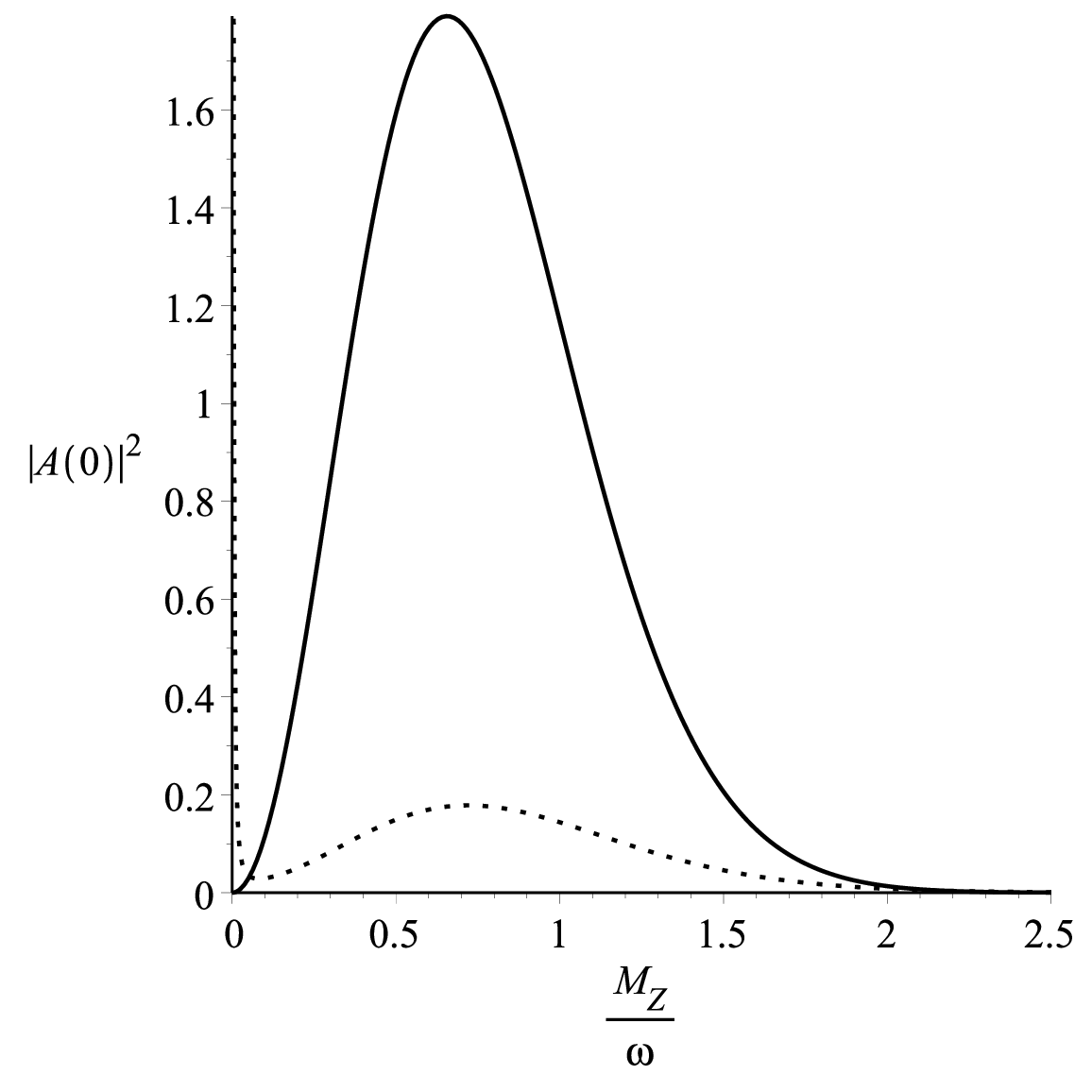}
\caption{$|B_{k}|^2$ as a function of parameter $M_Z/\omega$. Solid line is for $p=0.3,p'=0.6,\mathcal{P}=0.1$, while the point line is for $p=0.2,p'=0.6,\mathcal{P}=0.3 $ . $|A(0)|^2$ as a function of parameter $M_Z/\omega$. Solid line is for $p=0.3,p'=0.6,\mathcal{P}=0.1$, while the point line is for $p=0.2,p'=0.6,\mathcal{P}=0.3$ . }
\label{f1}
\end{figure}

\begin{figure}[h!t]
\includegraphics[scale=0.35]{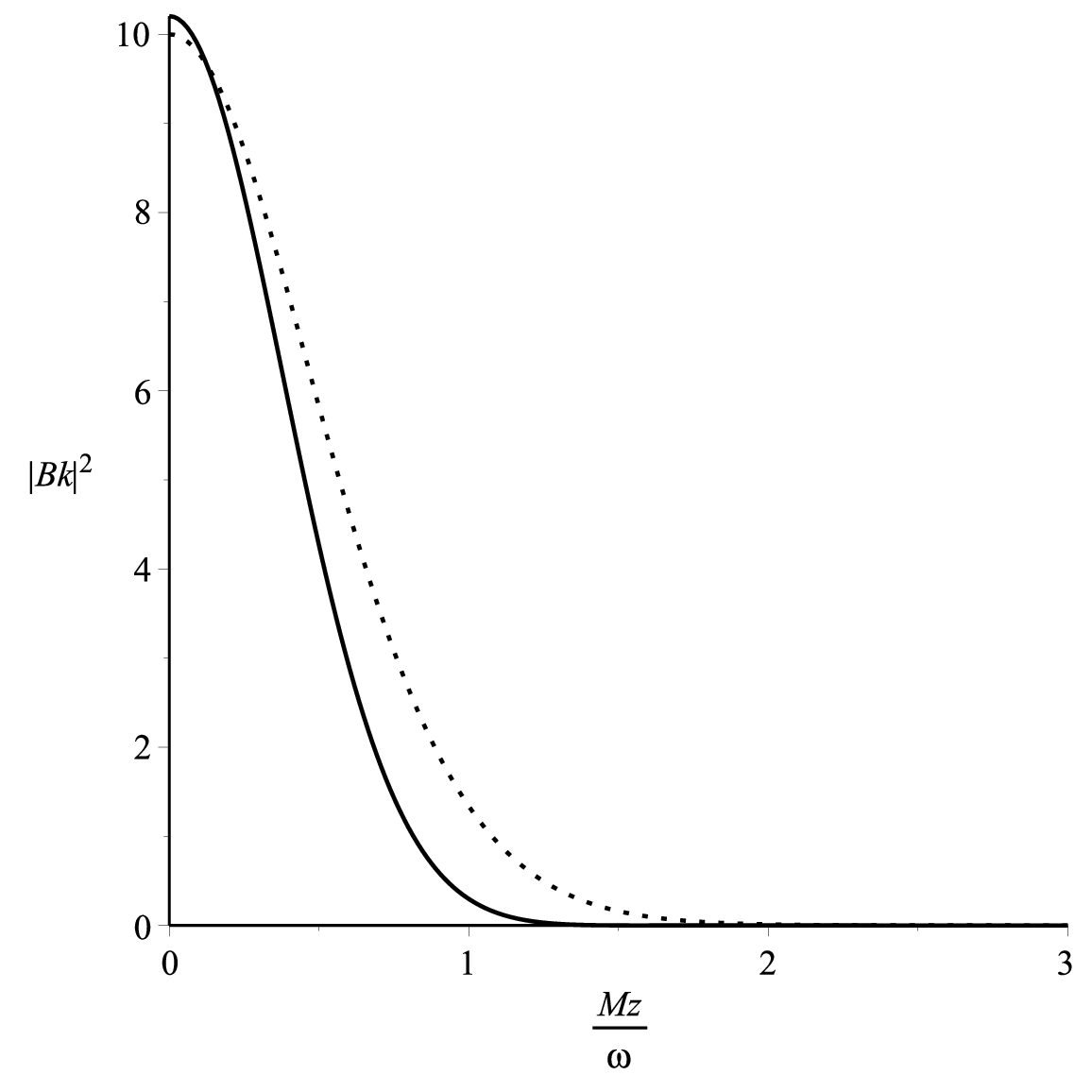}
\includegraphics[scale=0.35]{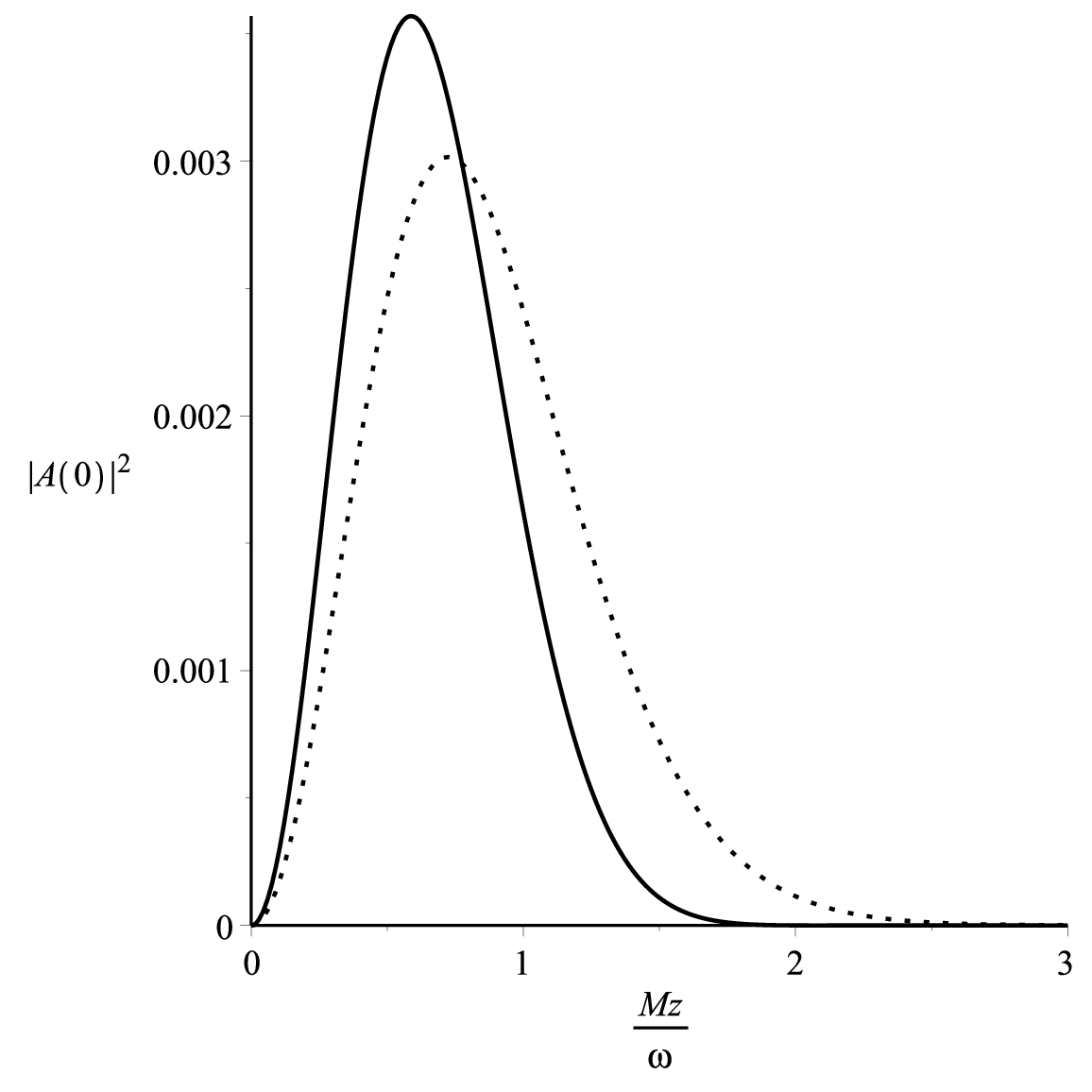}
\caption{$|B_{k}|^2$ as a function of parameter $M_Z/\omega$. Solid line is for $p=0.2,p'=0.8,\mathcal{P}=0.1$, while the dotted line is for $p=0.5,p'=0.8,\mathcal{P}=0.2 $ . $|A(0)|^2$ as a function of parameter $M_Z/\omega$. Solid line is for $p=0.2,p'=0.8,\mathcal{P}=0.1$, while the dotted line is for $p=0.5,p'=0.8,\mathcal{P}=0.2 $ . }
\label{f2}
\end{figure}

The final result is dependent on Gauss hypergeometric functions $_{2}F_{1}$ and gamma Euler functions $\Gamma$. The amplitudes depend on gravity via the parameter $k=\sqrt{\left(\frac{M_Z}{\omega}\right)^2-\frac{1}{4}}$. We also observe that the ratio between the Z boson mass and the expansion parameter $\frac{M_Z}{\omega}$, and the momenta $p,p',\mathcal{P}$ determine the analytical structure of the amplitudes.  The delta Dirac function $\delta^3(\vec{\mathcal{P}}+\vec{p}\,'-\vec{p}\,)$ ensures the momentum conservation in the process of Z boson emission by neutrino, and this factor will play a key role in the computations for obtaining the transition rate.
Because the amplitude is proportional with the delta Dirac function $\delta^3(\vec{\mathcal{P}}+\vec{p}\,\,'-\vec{p}\,)$, one can define the transition probability per volume unit, i.e. $|\delta^3(\vec{p}\,)|^2=V\delta^3(\vec{p}\,)$. For the production of Z bosons with $\lambda=0$ the probability is:
\begin{eqnarray}\label{pif}
&&P_{\nu\rightarrow Z\nu}(\lambda=0)=|\mathcal{A}_{\nu\rightarrow Z\nu}(\lambda=0)|^2=
\frac{e_0^2}{\sin^2(2\theta_{W})}\,\delta^3(\vec{\mathcal{P}}+\vec{p}\,'-\vec{p}\,)\frac{1}{(2\pi)^{3}}\nonumber\\
&&\left(\frac{1}{2}-\sigma\right)^2\left(\frac{1}{2}-\sigma'\right)^2\biggl\{
|A_{k}(\mathcal{P},p,p')|^2|\xi^+_{\sigma'}(\vec{p}\,')\vec{\sigma}\cdot\vec{\epsilon}\,^*(\vec{n}_{\mathcal{P}},\lambda=0)\xi_{\sigma}(\vec{p}\,)|^2\nonumber\\
&&+|C_{k}(\mathcal{P},p,p')|^2|\xi^+_{\sigma'}(\vec{p}\,')\xi_{\sigma}(\vec{p}\,)|^2\nonumber\\
&&+A_{k}^*(\mathcal{P},p,p')C_{k}(\mathcal{P},p,p')(\xi^+_{\sigma'}(\vec{p}\,')\vec{\sigma}\cdot\vec{\epsilon}\,^*(\vec{n}_{\mathcal{P}},\lambda=0)\xi_{\sigma}(\vec{p}\,))^*(\xi^+_{\sigma'}(\vec{p}\,')\xi_{\sigma}(\vec{p}\,))\nonumber\\
&&+C_{k}^*(\mathcal{P},p,p')A_{k}(\mathcal{P},p,p')(\xi^+_{\sigma'}(\vec{p}\,')\xi_{\sigma}(\vec{p}\,))^*(\xi^+_{\sigma'}(\vec{p}\,')\vec{\sigma}\cdot\vec{\epsilon}\,^*(\vec{n}_{\mathcal{P}},\lambda=0)\xi_{\sigma}(\vec{p}\,))\biggl\}.
\end{eqnarray}
The probability for the generation of transversal modes with $\lambda=\pm1$ is:
\begin{eqnarray}\label{pif1}
&&P_{\nu\rightarrow Z\nu}(\lambda=\pm1)=\frac{1}{2}\sum_{\lambda}|\mathcal{A}_{\nu\rightarrow Z\nu}(\lambda=\pm1)|^2=\frac{e_0^2}{\sin^2(2\theta_{W})}\,\delta^3(\vec{\mathcal{P}}+\vec{p}\,'-\vec{p}\,)\frac{1}{(2\pi)^{3}}\nonumber\\
&&\left(\frac{1}{2}-\sigma\right)^2\left(\frac{1}{2}-\sigma'\right)^2\biggl\{\frac{1}{2}\sum_{\lambda}|B_{k}(\mathcal{P},p,p')|^2|\xi^+_{\sigma'}(\vec{p}\,')\vec{\sigma}\cdot\vec{\epsilon}\,^*(\vec{n}_{\mathcal{P}},\,\lambda=\pm1)\xi_{\sigma}(\vec{p}\,)|^2\biggl\}.
\end{eqnarray}
Plotting the square modulus of the functions that define the probability for $\lambda=0,\pm1$, and taking the values on graphs from $M_Z/\omega=0$, we obtain the results from Fig. (\ref{f1}),(\ref{f2}).

Figures (\ref{f1}) and (\ref{f2}) show the variation of the probability density with the parameter $M_Z/\omega$ and prove that the process of Z boson emission is important when the Hubble parameter is larger than the mass of the boson. Another consequence of the graphs is that the probability is finite and nonvanishing for $\omega>>M_Z$ and also for $k=0$ or $M_Z/\omega=\frac{1}{2}$. The ratio used in our graphs must be understood as $\frac{M_Z\, c^2}{\hbar\,\omega}$ and we observe that the probability is nonvanishing as long as the energy of the background is larger or has the same order as the rest energy of the Z boson. This result corresponds to the well established knowledge that place the Z bosons production in the early universe.  Our graphical results for the probability prove that in the Minkowski limit $M_Z/\omega\rightarrow \infty$ the probabilities are vanishing and this result can also be obtained by taking the limit in the equations (\ref{pif}) and (\ref{pif1}). We can conclude that the process of Z boson emission by a neutrino is possible as a perturbative process only in early universe when the expansion factor is much larger than the Z boson mass.

\section{Total probability}
In this section we will analyse the situation when the expansion parameter $\omega$ is much larger than the Z boson mass. We will analyse the case of transversal polarization with $\lambda=\pm1$. This can be done by observing that in equation (\ref{abc1}) the algebraic argument of the Bessel K functions become very small when $t\rightarrow \infty$ and we can use the formula \cite{22}:
\begin{equation}\label{asim}
K_{\nu}(z)\simeq\frac{\Gamma(\nu)}{2}\left(\frac{2}{z}\right)^{\nu},\, z\rightarrow 0,
\end{equation}
where the real part of the index must meet the condition $Re(\nu)>0$.
In order to study the case where the expansion parameter is larger than the mass of the Z boson, we will consider the situation where $M_Z/\omega<<1/2$. We mention that both the probabilities and amplitudes have a good behaviour in terms of parameter of the $M_Z/\omega$ (see Figs. (\ref{f1}),(\ref{f2})). We compute the total probability of transition in the case of large expansion, when $M_Z/\omega\rightarrow 0$ and the index of Bessel K functions becomes $-ik\rightarrow\frac{1}{2}$.
The total probability is obtained by solving the integrals after the final momenta. These kind of integrals are in general divergent, and we will apply a method for regularization.
\begin{eqnarray}
  P_{tot} &=& \int d^3 P \int d^3 p' \, P_{i \rightarrow f}.
\end{eqnarray}
First we obtain the transition amplitude in the limit $M_Z/\omega \rightarrow 0$, by using equations (\ref{abc1}) and (\ref{cb}) :
\begin{eqnarray}
  A_{\nu \rightarrow Z \nu} (\lambda = \pm1) &=& \frac{e_0}{\sin (2\theta_W)} \frac{\delta^3 (\vec{P} + \vec{p}\,' - \vec{p})}{(2\pi)^{3/2}} \Big( \frac{1}{2} - \sigma\Big) \Big( \frac{1}{2} - \sigma' \Big) \frac{1}{\sqrt{2iP} (P+ p' - p)}  \nonumber  \\
  && \times \xi^+_{\sigma \,'} (\vec{p}\,') \vec{\sigma} \cdot \vec{\epsilon}^* \xi_{\sigma} (\vec{p}).
\end{eqnarray}
Then the total probability expression of Z boson emission by e neutrino in the limit of large expansion is
\begin{equation}
  P_{tot} = \frac{1}{2} \sum_{\lambda} \int d^3P \int d^3 p' \, \frac{e^2 \delta^3 (\vec{P} + \vec{p}\,' - \vec{p})}{2(2\pi)^3 \sin^2 (2\theta_W) P(P+p'-p)^2} \Big( \frac{1}{2} - \sigma \Big)^2 \Big( \frac{1}{2} - \sigma' \Big)^2 \big| \xi^+_{\sigma '} (\vec{p}\,') \vec{\sigma} \cdot \vec{\epsilon}^* \xi_{\sigma} (\vec{p}\,) \big|^2
\end{equation}
To facilitate our calculations we need to consider the case where the particles momenta are on the same direction, that is the $z$ axis, in such a way that the angle between the momenta vectors is $0$ i.e $\vec p=p\,\vec e_3\,,\,\vec p\,'=p'\,\vec e_3$. In this case the bispinor summation is reduced to a number. The momentum of the Z boson is also considered on the third axis such that $\vec P=P\,\vec e_3$. Then the momenta integrals in total probability expression are:
\begin{equation}
  I = \frac{1}{(2\pi)^3} \int d^3P \int d^3 p' \, \frac{\delta^3 (\vec{P} + \vec{p}\,' - \vec{p})}{P(P+p'-p)} = \frac{1}{(2\pi)^3} \int d^3P \, \frac{1}{2P(p-P)^2},
\end{equation}
where we perform the delta Dirac integration and finally we obtain an integral that is logarithmical divergent. To obtain a finite result we will apply the dimensional regularization proposed in \cite{41,42,43}. Then we will replace our integral with a $D$ dimensional momenta integral as follows:
\begin{equation}
  I(D) = \frac{1}{2} \int \frac{d^D P}{(2\pi)^D} \frac{1}{P(p-P)^2} = \frac{2\pi^{D/2}}{2(2\pi)^D \Gamma \big( \frac{D}{2} \big)} \int_{0}^{\infty} dP \, \frac{P^{D-1}}{P(p-P)^2}.
\end{equation}
The new variable $P=-py$ is introduced to obtain the integral of the Beta Euler function \cite{22}, as given in Appendix B
and the final result is:
\begin{eqnarray}
  I(D) &=& \frac{2\pi^{D/2}p^{D-3}}{2(2\pi)^D \Gamma \big( \frac{D}{2} \big) } \int_{0}^{\infty} \frac{dy \cdot y^{D-2}}{(1+y)^2} \nonumber\\
  &=& \frac{2\pi^{D/2}p^{D-3} \Gamma (D-1) \Gamma (3-D)}{2(2\pi)^D \Gamma \big( \frac{D}{2} \big) }.
\end{eqnarray}
Because the result is still divergent for $D=3$ we will use the relation between gamma Euler functions $z\Gamma(z)=\Gamma(1+z)$ to obtain
$\Gamma (3-D) =  \frac{\Gamma (4-D)}{(3-D)}$ ,that allows to rewrite $I(D) $ as:
\begin{equation}
  I(D) = \frac{2\pi^{D/2}p^{D-3} \Gamma (D-1) \Gamma (4-D)}{2(2\pi)^D \Gamma \big( \frac{D}{2} \big) (3-D)}.
\end{equation}
We observe that the above function is divergent in $D=3$, and to remove the divergence we use the method of minimal substraction proposed in \cite{43}. The pole of $I(D)$ in $D=3$ have the residue:
\begin{equation}
  ResI(D) = \lim_{D \rightarrow 3} (D-3) I(D) = -\frac{1}{4\pi^2}.
\end{equation}
The method of minimal substraction allows one to choose a counter-term dependent on a mass parameter $\mu$ with the general form $\frac{\mu^{s}R}{D-3}$, where $s$ is taken such the dimension of $ I(D)$ remains the same. Then the renormalized integral is:
\begin{eqnarray}
  I(D)_r &=& I(D) + \frac{\mu^{D-3}}{4\pi^2(D-3)} \\
  &=& \frac{1}{D-3} \bigg( \frac{2\pi^{D/2}p^{D-3} \Gamma (D-1) \Gamma (4-D)}{2(2\pi)^D \Gamma \big( \frac{D}{2} \big)} - \frac{\mu^{D-3}}{4\pi^2} \bigg).
\end{eqnarray}
The parenthesis from the above equation can be expanded around $D=3$, and the result is:
\begin{equation}
  \frac{2\pi^{D/2}p^{D-3} \Gamma (D-1) \Gamma (4-D)}{2(2\pi)^D \Gamma \big( \frac{D}{2} \big)} - \frac{\mu^{D-3}}{4\pi^2} = \frac{(D-3)}{8\pi^2} \bigg[2- \ln \Big( \frac{4\pi \mu^2}{p^2} \Big) - \psi \Big( \frac{3}{2} \Big) \bigg]+ \mathcal{O}((D-3)^2),
\end{equation}
where $\Psi(3/2)$ is the digamma Euler function.
Using this method we successfully cancel the divergent term $D-3$, and the final result of the renormalised integral is finite:
\begin{equation}
  I(D)_r = \frac{1}{8\pi^2} \bigg[2- \ln \Big( \frac{4\pi \mu^2}{p^2} \Big) - \psi \Big( \frac{3}{2} \Big) \bigg].
\end{equation}
The final expression for the total probability for the process of Z boson emission by a neutrino in the early universe depends on the fundamental constants of the electro-weak theory
\begin{equation}
  P_{tot} = \frac{e^2}{16 \pi^2 \sin^2 (2\theta_W)} \bigg[ 2-\ln \Big( \frac{4\pi \mu^2}{p^2} \Big) - \psi \Big( \frac{3}{2} \Big) \bigg]=\frac{\sqrt{2}M_W^2G_F \sin^2\theta_W}{4 \pi^2 \sin^2 (2\theta_W)} \bigg[2- \ln \Big( \frac{4\pi \mu^2}{p^2} \Big) - \psi \Big( \frac{3}{2} \Big) \bigg]
\end{equation}
where in the second equality we introduce the Fermi constant $G_F$ i.e. $e^2=4\sqrt{2}M_W^2G_F \sin^2\theta_W$, with $M_W$ the mass of the W boson.

\section{The rate of transition}
In the previous section we computed the total probability for the process of Z boson emission by a neutrino in de Sitter geometry, but a fundamental problem is related to the computation of measurable quantities for such a process. For that we will compute the rate of transition in this geometry, following the results obtained in \cite{cpc} for the Milne universe. Consider the transition amplitude between the initial and final state in a de Sitter space-time of the form:
\begin{equation}\label{dsrt}
A_{if}=\delta^3(\vec p_f-\vec p _i)M_{if}I_{if},
\end{equation}
where the delta functions assure the momentum conservation in the process. In equation (\ref{dsrt}) the usual delta function of energy $\delta(E_f-E_i)$, is missing in de Sitter amplitudes. This problem comes from the temporal integrals denoted by $I_{if}$:
\begin{equation}
I_{if}=\int_0^{\infty} dt \mathcal{K}_{if},
\end{equation}
whose results do not give the usual $\delta(E_f-E_i)$ as in Minkowski spacetime, but instead have a rather complex dependence on hypergeometric Gauss functions as in eqs.(\ref{l0}),(\ref{l1}), (\ref{abc1}). The rate of transition is defined in Minkowski space-time by using the fact that the four delta Dirac functions $\delta^4(p)$ when squared give the usual $\delta(0)\delta^3(0)=\frac{1}{(2\pi)^4}VT$, where $V$ is the volume and $T$ is the interaction time. Then the rate is obtained dividing the probability by $VT$, or in other words the rate is the probability derivative with respect to time. Since in de Sitter geometry this is no longer valid, we adopt the definition given in \cite{cpc}, where the derivative with respect to time is applied on the integrals $I_{if}$. These integrals are written in terms of conformal time $t_c$ and the transition rate will be defined in conformal chart $\{t_c,\vec x\}$ where we expect that the results are similar with those from Minkowski metric:
\begin{equation}
R_{if}=\lim_{t_c\rightarrow 0}\frac{1}{2V}\frac{d}{dt_c}|A_{if}|^2=\lim_{t\rightarrow \infty}\frac{e^{\omega t}}{2V}\frac{d}{dt}| A_{if}|^2.
\end{equation}
The above result for the rate of transition can be written now as:
\begin{equation}\label{rt}
R_{if}=\frac{1}{(2\pi)^3}\,\delta^3(\vec p_f-\vec p _i)| M_{if}|^2 | I_{if}| \lim_{t\rightarrow \infty}| e^{\omega t}\mathcal{K}_{if}|,
\end{equation}
where $\mathcal{K}_{if}$ are the integrands from the temporal integrals given in equations (\ref{abc1}). Since in the present paper the transition between states in continuum spectrum are discussed, then to obtain the rate we must integrate after the final momenta in equation (\ref{rt}). Then the rate definition for the process of Z emission by a neutrino is:
\begin{eqnarray}\label{rtfin}
R_{\nu\rightarrow Z\nu}=\frac{1}{(2\pi)^3}\int\frac{{d^3 p'}}{(2\pi)^3}\int \frac{d^3\mathcal{P}}{(2\pi)^3}\,\delta^3(\vec{\mathcal{P}}+\vec{p}\,'-\vec{p}\,)\sum_{\lambda}| M_{\nu\rightarrow Z\nu}(\lambda)|^2 | I_{\nu\rightarrow Z\nu}(\lambda)| \lim_{t\rightarrow \infty}| e^{\omega t}\mathcal{K}_{\nu\rightarrow Z\nu}(\lambda)|,
\nonumber\\
\end{eqnarray}
where $ I_{\nu\rightarrow Z\nu}$ are the temporal integrals given in equation (\ref{abc1}) with the results given in equations (\ref{ab}), (\ref{cb}). The quantity $ M_{\nu\rightarrow Z\nu}(\lambda)$, is proportional to the coupling constants and the polarization terms $\xi^+_{-\frac{1}{2}}(\vec{p}\,'\,)\vec{\sigma}\cdot\vec{\epsilon}\,^*(\vec{n}_{\mathcal{P}},\lambda)\xi_{-\frac{1}{2}}(\vec{p}\,)$. The function $\mathcal{K}_{\nu\rightarrow Z\nu}(\lambda)$ represents the integrands from equation (\ref{abc1}). All these quantities depend on $\lambda$. Knowing that the neutrino polarizations are fixed $\sigma=\sigma'=-\frac{1}{2}$, we need to sum only after the Z boson polarizations $\lambda$.

The transition rate can be evaluated in the limit where the expansion parameter is larger than the mass of the Z boson, by using the amplitude equation in this limit:

\begin{eqnarray}
  A_{\nu \rightarrow Z \nu} (\lambda = \pm1) &=& \frac{ie_0}{\sin (2\theta_W)} \frac{\delta^3 (\vec{P} + \vec{p}\,' - \vec{p}\,)}{(2\pi)^{3/2}} \Big( \frac{1}{2} - \sigma\Big) \Big( \frac{1}{2} - \sigma' \Big)   \nonumber  \\
  && \times \int_0^\infty dz \sqrt{z}\,e^{-i(p'-p)z}K_{1/2}(i\mathcal{P}z)\xi^+_{\sigma \,'} (\vec{p}\,') \vec{\sigma} \cdot \vec{\epsilon}^* \xi_{\sigma} (\vec{p}\,).
\end{eqnarray}
In the present case of large expansion $ I_{\nu\rightarrow Z\nu}$ has the expression:
\begin{equation}\label{inu}
 I_{\nu\rightarrow Z\nu}=i\int_0^\infty dz \sqrt{z}\,e^{-i(p'-p)z}K_{1/2}(i\mathcal{P}z)=\sqrt{\frac{\pi}{2i\mathcal{P}}}\frac{1}{p'+\mathcal{P}-p},
\end{equation}
while $M_{if}$ is given by:
\begin{equation}
M_{if}=\frac{e_0}{(2\pi)^{3/2}\sin (2\theta_W)}\,\xi^+_{\sigma \,'} (\vec{p}\,') \vec{\sigma} \cdot \vec{\epsilon}^* \xi_{\sigma} (\vec{p}\,).
\end{equation}
The limit from equation (\ref{rt}) can be obtained by using the integrand in equation (\ref{inu}):
\begin{equation}
  \lim_{t \to \infty} \Big|e^{\omega t} \mathcal{K}_{\nu\rightarrow Z\nu} \Big( \frac{M_Z}{\omega} = 0\Big) \Big| =  \frac{\sqrt{\pi}}{\sqrt{2 P}}.
\end{equation}

Putting all the above quantities together and summing after the polarization we obtain the transition rate:
\begin{equation}
  R_{i\rightarrow f} = \frac{1}{2} \sum_{\lambda} \frac{e_0^2 \, \delta^3 (\vec{P} + \vec{p}\,' - \vec{p}) \big| \xi^+_{\sigma '} (\vec{p}\,') \vec{\sigma} \cdot \vec{\epsilon}^* \xi_{\sigma} (\vec{p}\,) \big|^2}{(2\pi)^6 \sin^2 (2\theta_W) 2P (P+p'-p)},
\end{equation}
which can be simplified by choosing the momenta of particles on the same direction:
\begin{equation}
  R_{i\rightarrow f} = \frac{e_0^2}{\sin^2 (2\theta_W)} \frac{\delta^3 (\vec{P} + \vec{p}\,' - \vec{p}\,)}{(2\pi)^6 4P (P+p'-p)}
\end{equation}
The total transition rate is obtained by integration after the final momenta:
\begin{equation}
  R_{tot} = \int d^3P \int d^3 p' \, R_{i\rightarrow f} = \frac{1}{(2\pi)^3} \int d^3 p' \int d^3 P \frac{\delta^3 (\vec{P} + \vec{p}\,' - \vec{p}\,)}{P (P+p'-p)}.
\end{equation}
The momenta integral that needs to be computed is:
\begin{equation}
  I = \frac{1}{2(2\pi)^3} \int d^3 P \, \frac{1}{P(p-P)}.
\end{equation}
In this case we also use the dimensional regularization \cite{41,42,43} and the $D$ dimensional integral reads:
\begin{eqnarray}
  I(D) &=& \frac{1}{2(2\pi)^D} \int \frac{d^D P}{P(p-P)} = \frac{1}{2(2\pi)^D} \frac{2\pi^{D/2}}{\Gamma \big(\frac{D}{2} \big)} \int_{0}^{\infty} dP \, \frac{P^{D-1}}{P(p-P)} \nonumber \\
   &=& \frac{-2\pi^{D/2} p^{D-2} }{2(2\pi)^D \Gamma \big( \frac{D}{2} \big) } \int_{0}^{\infty} \frac{dy \cdot y^{D-2}}{(1+y)},
\end{eqnarray}
where the last equality in the above equation is obtained by making the variable change $P=-py$. The final result of the integral is obtained by using the definition of Beta Euler function as:
\begin{equation}
  I(D) = -\frac{2\pi^{D/2} p^{D-2}}{2(2\pi)^D \Gamma \big(\frac{D}{2} \big)} \frac{\Gamma(D-1) \Gamma(2-D)}{\Gamma(1)},
\end{equation}
which is divergent for $D=3$. To remove the divergence we apply the method of minimal substraction \cite{43}. First we rewrite the divergent gamma function as:
\begin{equation}
  \Gamma(2-D) = \frac{\Gamma(3-D)}{(2-D)} = \frac{\Gamma (4-D)}{(2-D)(3-D)}
\end{equation}
Then $ I(D)$ becomes
\begin{equation}
  I(D) = -\frac{2\pi^{D/2} p^{D-2} \Gamma (D-1)}{2(2\pi)^D \Gamma \big(\frac{D}{2} \big) } \frac{\Gamma (4-D)}{(2-D)(3-D)}
\end{equation}
The residue in $D=3$ is then computed
\begin{equation}
  ResI(D) = \lim_{D \rightarrow 3} (D-3) I(D) = - \frac{p}{4\pi^2}.
\end{equation}
Then the regularized integral is obtained by choosing a counter-term with the same dimension \cite{43}
\begin{eqnarray}
  I(D)_r &=& \frac{1}{(D-3)} \Bigg[ \frac{2\pi^{D/3} p^{D-2} \Gamma (D-1) \Gamma (4-D)}{(2\pi)^D  \Gamma \big(\frac{D}{2} \big) (2-D)} + \frac{p \mu^{D-3}}{4\pi^2} \Bigg] ,
\end{eqnarray}
where $\mu$ is a bosonic mass parameter.
The parenthesis expansion around $D=3$ then gives:
\begin{eqnarray}
 &&\Bigg[ \frac{2\pi^{D/3} p^{D-2} \Gamma (D-1) \Gamma (4-D)}{(2\pi)^D  \Gamma \big(\frac{D}{2} \big) (2-D)} + \frac{p \mu^{D-3}}{4\pi^2} \Bigg] \nonumber\\
  &\simeq& \bigg( \frac{p}{8\pi^2} \bigg) \bigg[ \psi \Big( \frac{3}{2} \Big) + \ln \Big( \frac{4\pi \mu^2}{p^2} \Big) \bigg] + \mathcal{O}\big((D-3)^2\big).
\end{eqnarray}
In this way one obtain the regularized result for the momenta integral:
\begin{equation}
  I(D)_r = \frac{p}{8\pi^2} \bigg[\psi \Big( \frac{3}{2}\Big) + \ln \Big( \frac{4\pi \mu^2}{p^2} \Big) \bigg].
\end{equation}
The total transition rate is then given by a finite expression:
\begin{equation}\label{fund}
  R_{tot} = \frac{e^2}{(2\pi)^3 \sin^2 (2\theta_W)} \frac{p}{32\pi^2} \bigg[ \psi \Big( \frac{3}{2}\Big) + \ln \Big( \frac{4\pi \mu^2}{p^2} \Big) \bigg]=
  \frac{\sqrt{2}M_W^2G_F \sin^2\theta_W}{(2\pi)^3 \sin^2 (2\theta_W)} \frac{p}{8\pi^2} \bigg[ \psi \Big( \frac{3}{2}\Big) + \ln \Big( \frac{4\pi \mu^2}{p^2} \Big) \bigg]
\end{equation}

In equation (\ref{fund}) we can obtain a numerical estimation for the transition rate of a Z boson emission by neutrino. Furthermore, one can establish the density number of Z bosons by taking into account the density numbers of neutrinos at different temperatures. For a very large number of neutrinos this process could have an important impact to the Z boson generation, as we will discuss in the following section.

\section{Density number of Z bosons}
In the process of Z boson emission by neutrinos the density number of neutrinos will determine the density number of Z bosons. For this reason we define the density number of Z bosons with our transition rate given in equation (\ref{fund}), as follows:
\begin{equation}
n_Z=\frac{R_{\nu\rightarrow Z\nu}}{R_d}\,n_{\nu}',
\end{equation}
where $R_d$ is the rate of Z boson decay and $n_{\nu}'$ is the density number of neutrinos in the conditions of early universe, with the observation that all three species will be taken into account. It is a well established fact that the Z bosons decay into fermion anti-fermion pairs and hadrons. The total rate of the decay for the Z boson is the sum of the mentioned rates and in Minkowski field theory its value is $R_d=2.5GeV$ \cite{37}.
For very early universe the temperature of photons and neutrinos are supposed to be equal $(T_{\gamma}=T_{\nu})$ such that the relations between the number of photons and neutrinos is \cite{36}:
\begin{equation}\label{nft}
n_{\nu}'=\frac{9}{4}n_{\gamma}'=\frac{9}{4}\frac{2\zeta(3)}{\pi^2}\left(\frac{k_B T'}{\hbar c}\right)^3,
\end{equation}
where the primed describe the density numbers in early universe.
From equations (\ref{nft}) it is clear that the density number of neutrinos in early universe will be higher, and we can say that the universe was dominated by neutrinos. As the temperature decreases and the universe expands, the neutrinos decouple and the number of neutrinos remains constant. This situation happens at temperatures around $10^{10} K$. Now let us comment the implications in relation with our rate given in equation (\ref{fund}).

Using the equations for decay rate of Z bosons (\ref{fund}) and number of neutrinos (\ref{nft}), the final equation for the density number of Z bosons produced in emission processes by neutrinos is:
\begin{equation}\label{nr}
n_Z=\frac{\sqrt{2}M_W^2G_F \sin^2\theta_W}{(2\pi)^3 \sin^2 (2\theta_W)R_d} \frac{p}{8\pi^2} \bigg[ \psi \Big( \frac{3}{2}\Big) + \ln \Big( \frac{4\pi \mu^2}{p^2} \Big) \bigg]\,n_{\nu}'
\end{equation}

Taking into account results given in equations (\ref{nft}) we can obtain the density number of Z bosons per cubic centimeter produced in emission processes by neutrinos. For example, we set the temperature at $T'=10^{14}K$ the density number of neutrinos given in equation (\ref{nft}) is $n_{\nu}'=4.56\cdot 10^{43}$, then the density number of Z bosons per cubic centimeter obtained from equation (\ref{nr}) is :
\begin{equation}
n_Z\simeq 7\cdot 10^{37} cm^{-3},
\end{equation}
where we take the ration $\frac{ \mu^2}{p^2}$ from the logarithm to be equal approximatively with one.
This result shows that in each cubic centimeter we have around $10^{37}$ Z bosons, when the temperature was $10^{14}K$. Using equation (\ref{nr}) one can obtain the number densities of Z bosons at different temperatures with the observation that for a realistic evaluation we need to keep the values around the temperatures specific to end of the electro-weak epoch and hadron epoch when the Z bosons become massive. Another observation is related to the fact that in our estimation we use the decay rate from Minkowski theory. This result can be improved by using the decay rates in de Sitter space-time that also need to be evaluated in the limit of large expansion factor.

At earlier times and high temperatures it is also possible to study the number of Z bosons with the observation that the mass of the Z boson could be smaller or close to zero in these conditions because the mass is related to the expectation values of the Higgs field.
Another important step for a better understanding of our results is related to the establishing of the variation of the Z boson mass with the temperature via the expansion parameter, or the dependence on the form $M_Z[T(\omega)]$, where $\omega$ is the Hubble constant. This is an important issue that we hope to analyse in a future research.

If the finite temperatures are considered then the forbidden processes from Minkowski theory are allowed because the energy received from the thermic bath, and this could be seen as similar to a certain degree to what the gravitational field can do on de Sitter space-time. This means that in the early universe the thermic effect and the expansion effect could compete as mechanisms that generate particle production. For this reason it is important to study both effects for a complete picture related to the problem of matter-antimatter generation in early universe.

In this paper we investigate the problem of particle production in early universe by using a perturbative method that accounts for the generation of particles at fields interactions. Our results prove that for a clear picture of the mechanisms that were involved in matter production in early universe one should take into account the perturbative mechanism, based on computations of the first order transition amplitudes that have nonvanishing contributions in a non-stationary geometry. The mechanism proposed here could be one of the possible explanations for the abundance of Z bosons in the conditions of the early universe.
We must point out that the rate computed in our paper is valid in the conformal chart $\{t_c,\vec x\}$ and the result was obtained by using the modes for the Dirac field and Proca field which are defined globally in de Sitter manifold. Thus we do not obtain the rate dependence on the observer since this will imply the using of different modes in different charts and the Bogoliubov transformations. However the perturbative result for the rate should be considered along with the cosmological results for a complete picture about the problem of particle production in early universe as was proved in \cite{19}. To the best of our knowledge the problem of cosmological production of massive bosons is less studied in literature and we hope that our perturbative results will be the start for the study of the Bogoliubov transformations in the case of Proca field, that should also give the nonperturbative density number of  massive bosons. One of the notable results related to the production of massive vector bosons in de FRLW universe was obtained in \cite{40}, where the density number was obtained using the Bogoliubov transformations. For futher studies it will be interesting to establish the behaviour of the transition rate (\ref{rtfin}) and density number (\ref{nr}) at isometry transformations on de Sitter space-time. This should establish the invariance of this quantities in de Sitter geometry.

\section{Appendix A: Free fields in de Sitter geometry}
In equation (\ref{am2}) the solutions of the Dirac equations for zero mass field, $(U_{p\sigma})_{\nu}(x)$ are used. These solutions describe the neutrino field in de Sitter geometry and their explicit form was obtained in \cite{3}:
\begin{eqnarray}\label{sol1}
(U_{\vec{p},\sigma}(x))_{\nu}=\left(-\frac{\omega t_{c}}{2\pi}\right)^{3/2}\left (\begin{array}{c}
(\frac{1}{2}-\sigma) \xi_{\sigma}(\vec{p}\,)\\
0
\end{array}\right)e^{i\vec{p}\cdot\vec{x}-i p t_{c}}.
\end{eqnarray}
The form of the helicity bispinors can be expressed as follows \cite{14}:
\begin{equation}\label{spin}
\xi_{\frac{1}{2}}(\vec{p}\,)=\sqrt{\frac{p_3+p}{2 p}}\left(
\begin{array}{c}
1\\
\frac{p_1+ip_2}{p_3+p}
\end{array} \right)\,,\quad
\xi_{-\frac{1}{2}}(\vec{p}\,)=\sqrt{\frac{p_3+p}{2 p}}\left(
\begin{array}{c}
\frac{-p_1+ip_2}{p_3+p}\\
1
\end{array} \right)\,,
\end{equation}
while $\eta_{\sigma}(\vec{p}\,)= i\sigma_2
[\xi_{\sigma}(\vec{p}\,)]^{*}$. These spinors satisfy the relation:
\begin{equation}\label{pa}
\vec{\sigma}\vec{p}\,\xi_{\sigma}(\vec{p}\,)=2p\,\sigma\xi_{\sigma}(\vec{p}\,)
\end{equation}
with $\sigma=\pm\frac{1}{2}$, where $\vec{\sigma}$ are the Pauli
matrices and $p=\mid\vec{p}\mid$ is the modulus of the momentum vector.
In the case of Proca field the temporal and spatial solutions in de Sitter geometry are obtained in \cite{4}.
The spatial part of the solution is given by:
\begin{eqnarray}\label{sol2}
\vec{f}_{\vec{\mathcal{P}},\lambda}(x)=\left\{
\begin{array}{cll}
\frac{i\sqrt{\pi}\omega \mathcal{P}e^{-\pi k/2}}{2M_Z(2\pi)^{3/2}}\left[(\frac{1}{2}+ik)\frac{\sqrt{-t_{c}}}{\mathcal{P}}
H^{(1)}_{ik}\left(-\mathcal{P}t_{c}\right)-(-t_{c})^{3/2}H^{(1)}_{1+ik}\left(-\mathcal{P}t_{c}\right)\right]e^{i\vec{\mathcal{P}}\vec{x}}\vec{\epsilon}\,(\vec{n}_{\mathcal{P}},\lambda)&{\rm for}&\lambda=0\\
\frac{\sqrt{\pi}e^{-\pi k/2}}{2(2\pi)^{3/2}}\sqrt{-t_{c}}H^{(1)}_{ik}\left(-\mathcal{P}t_{c}\right)e^{i\vec{\mathcal{P}}\vec{x}}\vec{\epsilon}\,(\vec{n}_{\mathcal{P}},\lambda)
&{\rm for}&\lambda=\pm 1.
\end{array}\right.
\nonumber\\
\end{eqnarray}

while the temporal part of the solution of the Proca equation \cite{4} is given by:
\begin{eqnarray}\label{fo}
f_{0\vec{\mathcal{P}},\lambda}(x)=\left\{
\begin{array}{cll}
\frac{\sqrt{\pi}\omega\mathcal{P}e^{-\pi k/2}}{2M_Z(2\pi)^{3/2}}(-t_{c})^{3/2}H^{(1)}_{ik}\left(-\mathcal{P}t_{c}\right)e^{i\vec{\mathcal{P}}\vec{x}}&{\rm for}&\lambda=0\\
0
&{\rm for}&\lambda=\pm 1.
\end{array}\right.
\end{eqnarray}

In the solutions for the Proca equation in momentum basis the polarization vectors are $\vec{n}_{\mathcal{P}}=\vec{\mathcal{P}}/\mathcal{P}$ and $\vec{\epsilon}\,(\vec{n}_{\mathcal{P}},\lambda)$. For $\lambda=\pm 1$ the polarization vectors are transversal,
$\vec{\mathcal{P}}\cdot\vec{\epsilon}\,(\vec{n}_{\mathcal{P}},\lambda=\pm1)=0$. In the case $\lambda=0$ the polarization vectors are longitudinal on the momentum
$\vec{\mathcal{P}}\cdot\vec{\epsilon}\,(\vec{n}_{\mathcal{P}},\lambda=0)=\mathcal{P}$, since $\vec{\epsilon}\,(\vec{n}_{\mathcal{P},\lambda=0})=\vec{n}_{\mathcal{P}}$. The notation for the mass of the Z boson is $M_Z$, and the parameter $k=\sqrt{\left(\frac{M_Z}{\omega}\right)^2-\frac{1}{4}}$ is dependent on the ratio $\frac{M_Z}{\omega}$, and the condition  $\frac{M_Z}{\omega}>\frac{1}{2}$ assures that the index of the Hankel function is imaginary.

\section{Appendix B: Integrals for obtaining the amplitude and rate}

The temporal integrals from amplitude are of the type \cite{22}:
\begin{eqnarray}\label{a3}
&&\int_0^{\infty} dz
z^{\mu-1}e^{-\alpha z}K_{\nu}(\beta z)=\frac{\sqrt{\pi}(2\beta)^{\nu}}{(\alpha+\beta)^{\mu+\nu}}\frac{\Gamma\left(\mu+\nu\right)\Gamma\left(\mu-\nu\right)}{\Gamma\left(\mu+\frac{1}{2}\right)}\nonumber\\
&&\times\,_{2}F_{1}\left(\mu+\nu,\nu+\frac{1}{2};\mu+\frac{1}{2};\frac{\alpha-\beta}{\alpha+\beta}\right),\nonumber\\
&&Re(\alpha+\beta)>0\,,|Re(\mu)|>|Re(\nu)|.
\end{eqnarray}

The integral of the Beta Euler function is given by:
\begin{eqnarray}
B(a,b)=\frac{\Gamma(a)\Gamma(b)}{\Gamma(a+b)}=\int_0^{\infty}dy \frac{y^{a-1}}{(1+y)^{a+b}}
\end{eqnarray}
and in our case the values for arguments are $a=D-1\,,b=3-D$ .

\textbf{Acknowledgements}
This work was supported by a grant of the Romanian Ministry of Research and Innovation and West University of Timi\c soara, CCCDI-UEFISCDI, under project "VESS, 18PCCDI/2018",  within PNCDI III.

We would like to thank dr. Victor Ambrus for suggestions to improve the manuscript.

\end{document}